\documentclass[preprint]{aastex}
\usepackage{bm}       % bold math
\begin{document}
%%%%%%%%%%%%%%%%%%%%%%%%%%%%%%%%%%%%%%%%%%%%%%%%%%%%%%%%%%%%%%%%%%
\title{%
 MEASURING PRIMORDIAL NON-GAUSSIANITY IN THE COSMIC MICROWAVE BACKGROUND
}%

\author{%
 E. Komatsu\altaffilmark{1},
 D. N. Spergel\altaffilmark{2},
 B. D. Wandelt\altaffilmark{3}
}%

\altaffiltext{1}{%
 Department of Astronomy, University of Texas at Austin,
 Austin, TX 78712
}%

\altaffiltext{2}{%
 Department of Astrophysical Sciences, Princeton University,
 Princeton, NJ 08544
}%

\altaffiltext{2}{%
  Department of Physics, University of Illinois at Urbana-Champaign,
  Urbana, IL 61801-3080
}%

%\email{komatsu@astro.princeton.edu}
\email{komatsu@astro.as.utexas.edu}
\date{\today}
%%%%%%%%%%%%%%%%%%%%%%%%%%%%%%%%%%%%%%%%%%%%%%%%%%%%%%%%%%%%%%%%%%
\begin{abstract}
%1-------10--------20--------30--------40--------50--------60%
 We derive a fast way for measuring primordial non-Gaussianity 
  in a nearly full-sky map of the cosmic microwave background.
  We find a cubic combination of sky maps combining bispectrum
  configurations to capture a quadratic term in primordial   
  fluctuations. Our method takes only $N^{3/2}$ operations rather than 
  $N^{5/2}$ of the bispectrum analysis (1000 times faster for $l=512$), 
  retaining the same sensitivity.
  A key component is a map of underlying primordial fluctuations, which 
  can be more sensitive to the primordial non-Gaussianity than a 
  temperature map. We also derive a fast and accurate 
  statistic for measuring non-Gaussian signals from foreground 
  point sources. The statistic is $10^6$ times faster than the 
  full bispectrum analysis, and can be used to estimate contamination
  from the sources. Our algorithm has been successfully applied to
  the {\sl Wilkinson Microwave Anisotropy Probe} sky maps by
  \citet{2003astro.ph..2223K}.
%1-------10--------20--------30--------40--------50--------60% 
\end{abstract}
%%%%%%%%%%%%%%%%%%%%%%%%%%%%%%%%%%%%%%%%%%%%%%%%%%%%%%%%%%%%%%%%%%
% \pacs{98.70.Vc,98.80.-k,98.80.Es}
% \maketitle
%%%%%%%%%%%%%%%%%%%%%%%%%%%%%%%%%%%%%%%%%%%%%%%%%%%%%%%%%%%%%%%%%%
%
% Introduction
%
%%%%%%%%%%%%%%%%%%%%%%%%%%%%%%%%%%%%%%%%%%%%%%%%%%%%%%%%%%%%%%%%%%
\section{INTRODUCTION}\label{sec:intro}

Measurement of statistical properties of the cosmic microwave background
(CMB) is a direct test of inflation.
Simple models of inflation predict Gaussian primordial
fluctuations generated by ground-state quantum fluctuations of 
a scalar field
\citep{1982PhRvL..49.1110G,Starobinsky:1982ee,Hawking:1982cz,1983PhRvD..28..679B,Mukhanov:1992me}.

Non-linearity \citep{1990PhRvD..42.3936S,1991PhRvD..43.1005S,1994ApJ...430..447G,Gupta:2002kn},
interactions of scalar fields \citep{Allen:1987vq,1993ApJ...403L...1F},
or deviation from the ground state
\citep{Lesgourgues:1997jc,Martin:1999fa,Contaldi:1999jr,2002astro.ph..5202G} 
can generate weak non-Gaussianity.
\citet{Acquaviva:2002ud} and \citet{Maldacena:2002vr} have calculated the 
2nd-order perturbations during inflation to show that simple inflation 
based on a slowly rolling scalar field cannot generate 
detectable non-Gaussianity with the 
{\sl Wilkinson Microwave Anisotropy Probe} ({\sl WMAP}) or 
the {\sl Planck} experiments; thus, any detection of the primordial 
non-Gaussianity strongly constrains inflation models, and sheds light 
on physics in the early universe. For example, isocurvature fluctuations
\citep{Linde:1997gt,1997ApJ...483L...1P,Bucher:1997gg},
features in a scalar-field potential 
\citep{1991lssp.conf..339K,Wang:1999vf}, or a ``curvaton'' mechanism
(by which late-time decay of a scalar field generates
curvature perturbations from isocurvature fluctuations \citep{Mollerach:1990hu,Lyth:2001nq}) 
can generate stronger, potentially detectable, non-Gaussianity.
The second-order gravity also contributes to non-Gaussianity 
\citep{Luo:1993ei,1995ApJ...454..552M,Pyne:1996bs,Mollerach:1997up},
and there is a possibility that we can detect it with the {\sl Planck} 
experiment.

%
% Model
%
Many of the non-Gaussian models are written as 
a parametrized form for the curvature perturbations
$\Phi$,
%%%%%%%%%%%%%%%%%%%%%%%%%%%%%%%%%%%%%%%%%%%%%%%%%%%%%%%%%%%%%%%%%%
\begin{equation}
 \label{eq:phi}
 \Phi({\bm x})= \Phi_L({\bm x}) + f_{NL}\left[ \Phi_L^2({\bm x}) - \left<\Phi_L^2({\bm x})\right> \right],
\end{equation}
%%%%%%%%%%%%%%%%%%%%%%%%%%%%%%%%%%%%%%%%%%%%%%%%%%%%%%%%%%%%%%%%%%
where $\Phi_L$ are Gaussian linear perturbations.
Note that $\Phi=\Phi_H$ in \citet{Bardeen:1980kt}.
A similar model may also apply to isocurvature perturbations, 
$S$ \citep{Bartolo:2001cw}.
This \textit{ansatz} provides a model quantifying the amplitude
of the primordial non-Gaussianity. 
An exact prediction of this model for the CMB bispectrum 
exists \citep{2001PhRvD..63f3002K}, while an approximate one 
for the trispectrum \citep{Okamoto:2002ik}.
Non-linearity in inflation gives 
$f_{NL}\sim {\cal O}(10^{-1})$, the second-order gravity gives 
$f_{NL}\ga {\cal O}(1)$, and isocurvature fluctuations, features, 
or curvatons can give $f_{NL}\gg 1$ depending on models.

%
% How large can f_NL be? observational constraints
%
The bispectrum measured by {\sl COBE} \citep{2002ApJ...566...19K} and 
{\sl MAXIMA} \citep{2002astro.ph.11123S} experiments found 
$\left|f_{NL}\right|\la 10^3$ (68\%).
Using the methods described in this paper, the {\sl WMAP} data 
\citep{2003astro.ph..2207B} have improved the constraint significantly 
to obtain $-58<f_{NL}<134$ (95\%) \citep{2003astro.ph..2223K}.
While the trispectrum measured on the {\sl COBE} data has shown no evidence 
for cosmological non-Gaussianity 
\citep{2002astro.ph..6039K,2001ApJ...563L..99K}, no quantitative
limit on $f_{NL}$ has been obtained.
The trispectrum can be as sensitive to $f_{NL}$ as the bispectrum 
\citep{Okamoto:2002ik}; however, we need more accurate predictions
using the full radiation transfer function.
The trispectrum has not yet been measured on the {\sl WMAP} data.
The deficit in $C^2(\theta)$ on $\theta\ga 60^\circ$ 
\citep{2003astro.ph..2209S} might be a sign of
a significant trispectrum on large angular scales.

%
% Why need a fast method?
%
Measuring $f_{NL}$ from nearly full-sky experiments is challenging. 
The bispectrum analysis requires $N^{5/2}$ operations 
($N^{3/2}$ for computing three $l$'s and $N$ for averaging over the sky) 
where $N$ is the number of pixels
($N\sim 3\times 10^6$ for {\sl WMAP}, $5\times 10^7$ for {\sl Planck}).
Even though an efficient algorithm exists, 
the trispectrum still requires $N^3$ \citep{2002astro.ph..6039K}.

Although we measure the individual triangle configurations of the bispectrum
(or quadrilateral configurations of the trispectrum) at first, 
we eventually combine all of them to constrain model parameters
such as $f_{NL}$, as the signal-to-noise per configuration is nearly zero.
This may sound inefficient. Measuring all configurations 
is enormously time consuming. 
Is there any statistic which \textit{already} 
combines all the configurations optimally, and fast to compute?
Yes, and finding it is the main subject of this paper.
A physical justification for our methodology is as follows.
A model like equation~(\ref{eq:phi}) generates non-Gaussianity in real 
space, and central-limit theorem makes the Fourier modes nearly 
Gaussian; thus, real-space statistics should be more sensitive.
On the other hand, real-space statistics are weighted sum of Fourier-space 
statistics, which are often easier to predict.
Therefore, we need to understand the shape of Fourier-space statistics
to find sensitive real-space statistics, and for this purpose it
is useful to have a specific, physically motivated non-Gaussian 
model, compute Fourier statistics, and find optimal real-space
statistics.

%
% a_lm <--> primordial fluctuations
%
%%%%%%%%%%%%%%%%%%%%%%%%%%%%%%%%%%%%%%%%%%%%%%%%%%%%%%%%%%%%%%%%%%
\section{RECONSTRUCTING PRIMORDIAL FLUCTUATIONS FROM TEMPERATURE ANISOTROPY}\label{sec:wiener}

We begin with the primordial curvature perturbations
$\Phi\left({\bm x}\right)$ and isocurvature perturbations
$S\left({\bm x}\right)$.
If we can reconstruct these primordial fluctuations from 
observed CMB anisotropy, $\Delta T(\hat{\bm n})/T$, then we can 
improve sensitivity to primordial non-Gaussianity. 
We find that the harmonic coefficients of CMB anisotropy, 
$a_{lm}=
T^{-1}\int d^2\hat{\mathbf n}\Delta T(\hat{\bm n}) Y_{lm}^*(\hat{\bm n})$,
are related to the primordial fluctuations as
%%%%%%%%%%%%%%%%%%%%%%%%%%%%%%%%%%%%%%%%%%%%%%%%%%%%%%%%%%%%%%%%%%
\begin{equation}
 \label{eq:alm}
  a_{lm}= b_l\int r^2 dr 
  \left[ \Phi_{lm}(r)\alpha_l^{adi}(r)
       + S_{lm}(r)\alpha_l^{iso}(r) \right] + n_{lm},
\end{equation}
%%%%%%%%%%%%%%%%%%%%%%%%%%%%%%%%%%%%%%%%%%%%%%%%%%%%%%%%%%%%%%%%%%
where $\Phi_{lm}(r)$ and $S_{lm}(r)$ are the harmonic coefficients
of the fluctuations at a given comoving distance, $r=\left|{\bm x}\right|$. 
A beam function $b_l$ and the harmonic coefficients of noise 
$n_{lm}$ represent instrumental effects.
Since noise can be spatially inhomogeneous, the noise covariance matrix
$\left<n_{lm}n_{l'm'}^*\right>$ can be non-diagonal; however, we approximate
it with $\simeq \sigma_0^2\delta_{ll'}\delta_{mm'}$. 
We thus assume the ``mildly inhomogeneous'' noise for which 
this approximation holds.
The function $\alpha_l(r)$ is defined by
%%%%%%%%%%%%%%%%%%%%%%%%%%%%%%%%%%%%%%%%%%%%%%%%%%%%%%%%%%%%%%%%%%%
\begin{equation}
 \label{eq:alpha_l}
 \alpha_l(r)
  \equiv
  \frac{2}{\pi}\int k^2 dk g_{Tl}(k) j_l(k r),
\end{equation}
%%%%%%%%%%%%%%%%%%%%%%%%%%%%%%%%%%%%%%%%%%%%%%%%%%%%%%%%%%%%%%%%%%%
where $g_{Tl}(k)$ is the radiation transfer function of either
adiabatic ($adi$) or isocurvature ($iso$) perturbations.
% Note that this function constitutes the primordial CMB bispectrum
% ($f_{NL}^{-1}b_l^{NL}(r)$ in \citep{2001PhRvD..63f3002K}). 

%
% Map of primordial fluctuations
%
Next, assuming that $\Phi\left({\bm x}\right)$ dominates,
we try to reconstruct $\Phi\left({\bm x}\right)$ from
the observed $\Delta T(\hat{\bm n})$.
A linear filter, ${\cal O}_l(r)$, which reconstructs
the underlying field, can be obtained by minimizing variance of 
difference between the filtered field ${\cal O}_l(r) a_{lm}$ and
the underlying field $\Phi_{lm}(r)$.
By evaluating 
%%%%%%%%%%%%%%%%%%%%%%%%%%%%%%%%%%%%%%%%%%%%%%%%%%%%%%%%%%%%%%%%%%%
\begin{equation}
 \label{eq:minimization}
  \frac{\partial}{\partial {\cal O}_l(r)}
  \langle\left|{\cal O}_l(r) a_{lm}-\Phi_{lm}(r)\right|^2\rangle
  =0,
\end{equation}
%%%%%%%%%%%%%%%%%%%%%%%%%%%%%%%%%%%%%%%%%%%%%%%%%%%%%%%%%%%%%%%%%%%
one obtains a solution for the filter as
%%%%%%%%%%%%%%%%%%%%%%%%%%%%%%%%%%%%%%%%%%%%%%%%%%%%%%%%%%%%%%%%%%%
\begin{equation}
 \label{eq:Ol}
  {\cal O}_l(r)= \frac{\beta_l(r)b_l}{{\cal C}_l},
\end{equation}
%%%%%%%%%%%%%%%%%%%%%%%%%%%%%%%%%%%%%%%%%%%%%%%%%%%%%%%%%%%%%%%%%%%
where the function $\beta_l(r)$ is given by
%%%%%%%%%%%%%%%%%%%%%%%%%%%%%%%%%%%%%%%%%%%%%%%%%%%%%%%%%%%%%%%%%%%
\begin{equation}
 \label{eq:beta_l}
  \beta_l(r) \equiv 
  \frac{2}{\pi}\int k^2 dk P(k) g_{Tl}(k) j_l(k r),
\end{equation}
%%%%%%%%%%%%%%%%%%%%%%%%%%%%%%%%%%%%%%%%%%%%%%%%%%%%%%%%%%%%%%%%%%%
and $P(k)$ is the power spectrum of $\Phi$.
Of course, one can replace $\Phi$ with $S$ when $S$ dominates.
% This function also constitutes the primordial bispectrum
% ($b_l^L(r)$ in \citep{2001PhRvD..63f3002K}).
% Here $\beta_l(r)$ is $b_l^L(r)$ in \citep{2001PhRvD..63f3002K}, which 
% as well as $\alpha_l(r)$.
We use a calligraphic letter for a quantity that includes effects
of $b_l$ and noise such that 
%%%%%%%%%%%%%%%%%%%%%%%%%%%%%%%%%%%%%%%%%%%%%%%%%%%%%%%%%%%%%%%%%%%
% \begin{equation}
${\cal C}_l \equiv C_lb_l^2+\sigma_0^2$,
% \end{equation}
%%%%%%%%%%%%%%%%%%%%%%%%%%%%%%%%%%%%%%%%%%%%%%%%%%%%%%%%%%%%%%%%%%%
where $C_l$ is the theoretical power spectrum that uses the same cosmological
model as $g_{Tl}(k)$.

Finally, we transform the filtered field ${\cal O}_l(r) a_{lm}$ back to 
pixel space to obtain an Wiener-filtered, reconstructed map of 
$\Phi(r,\hat{\bm n})$ or $S(r,\hat{\bm n})$.
We have assumed that there is no correlation between $\Phi$ and $S$.
We will return to study the case of non-zero correlation later
(\S~\ref{sec:mix}).

\begin{figure}
% \leavevmode \epsfxsize=8.5cm \epsfbox{figure1.eps}
% \includegraphics{figure1.eps}% Here is how to import EPS art
 \plotone{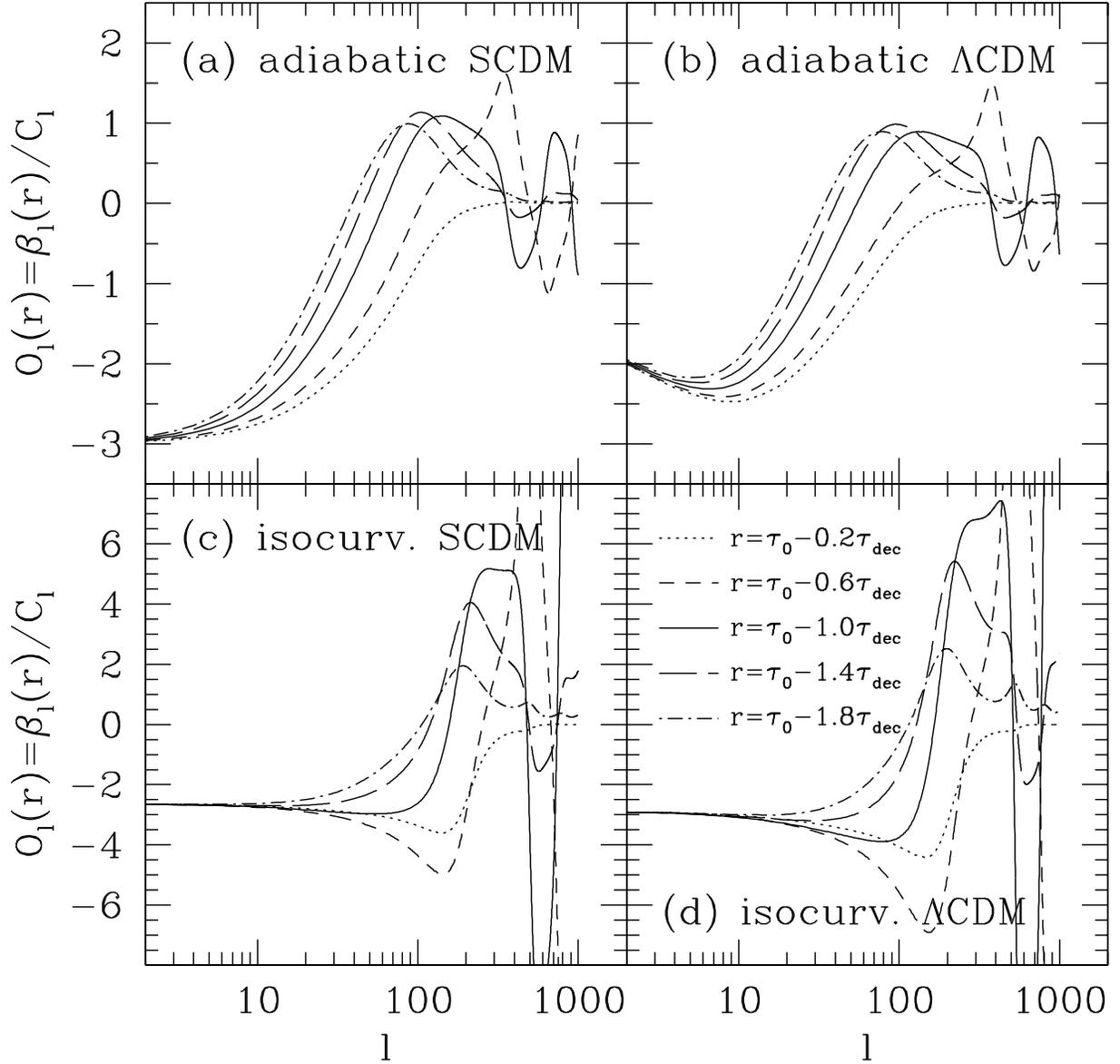}
 \caption{\label{fig:filters} 
 Wiener filters for the primordial fluctuations applied to 
 a CMB sky map, $O_l(r)=\beta_l(r)/C_l$ [Eq.~(\ref{eq:Ol})].
 We plot (a) $O_l$ for an adiabatic SCDM 
 ($\Omega_m=1$, $\Omega_\Lambda=0$, $\Omega_b=0.05$, $h=0.5$),
 (b) for an adiabatic $\Lambda$CDM ($\Omega_m=0.3$, 
 $\Omega_\Lambda=0.7$, $\Omega_b=0.04$, $h=0.7$), 
 (c) for an isocurvature SCDM, and
 (d) for an isocurvature $\Lambda$CDM.
 The filters are plotted at five conformal distances 
 $r=c(\tau_0-\tau)$ as explained in the bottom-right panel.
 Here $\tau$ is the conformal time ($\tau_0$ at the present).
 The SCDM models have $c\tau_0=11.84$~Gpc and 
 $c\tau_{dec}=0.235$~Gpc, while the $\Lambda$CDM models
 $c\tau_0=13.89$~Gpc and $c\tau_{dec}=0.277$~Gpc,
 where $\tau_{dec}$ is the photon decoupling epoch.}
\end{figure}

%
% Show Wiener filters
%
Figure~\ref{fig:filters} shows $O_l(r)$ as a function of $l$ and 
$r$ for 
(a) an adiabatic SCDM ($\Omega_m=1$), 
(b) an adiabatic $\Lambda$CDM ($\Omega_m=0.3$), 
(c) an isocurvature SCDM, and
(d) an isocurvature $\Lambda$CDM.
Note that we plot $O_l(r)=\beta_l(r)/C_l$ where $C_l$ does not
include beam smearing or noise.
While we have used $P(k)\propto k^{-3}$ for both adiabatic and
isocurvature modes, specific choice of $P(k)$ does not affect 
$O_l$ very much as $P(k)$ in $\beta_l$ in the numerator approximately
cancels out $P(k)$ in $C_l$ in the denominator. 
On large angular scales (smaller $l$) the Sachs--Wolfe (SW) effect
makes $O_l$ $-3$ for adiabatic modes and $-5/2$ for 
isocurvature modes of the SCDM models \citep{1967ApJ...147...73S}.
For the $\Lambda$CDM models the late-time decay of gravitational potential
makes this limit different. 
Adiabatic and isocurvature modes are out of phases in $l$.

%%%%%%%%%%%%%%%%%%%%%%%%%%%%%%%%%%%%%%%%%%%%%%%%%%%%%%%%%%%%%%%%%%
%%%%%%%%%%%%%%%%%%%%%%%%%%%%%%%%%%%%%%%%%%%%%%%%%%%%%%%%%%%%%%%%%%

%
% Why does the cubic statistic work?
%
The figure shows that $O_l$ changes the sign of the fluctuations as a 
function of scales.
This indicates that acoustic physics at the last scattering surface
modulates fluctuations so that hot spots in the primordial 
fluctuations can be cold spots in CMB for example.
Therefore, the shape of $O_l$ ``deconvolves'' the sign change,
recovering the phases of fluctuations.
This is an intuitive reason why our cubic statistic derived below
[Eq.~(\ref{eq:skewness})] works, and it proves more advantageous 
to measure primordial non-Gaussianity on a filtered map than on a 
temperature map.

This property should be compared to that of real-space statistics 
measured on a temperature map. 
We have shown in \citet{2001PhRvD..63f3002K} that the skewness of 
a temperature map is much less sensitive to the primordial 
non-Gaussianity than the bispectrum, exactly because of the cancellation 
effect from the acoustic oscillations. 
The skewness of a filtered map, on the other hand,
has a larger signal-to-noise ratio, and a more optimal statistic like our 
cubic statistic derived below (\S~\ref{sec:cubic}) can be constructed.
Other real-space statistics such as Minkowski functionals or peak-peak 
correlations may also be more sensitive 
to the primordial non-Gaussianity, when measured on the filtered 
maps; we are investigating these possibilities.

%
% Unavoidable loss of information?
%
Unfortunately, as $g_{Tl}$ oscillates, our reconstruction of $\Phi$ or 
$S$ from a temperature map alone is not perfect.
While $O_l$ reconstructs the primordial fluctuations very well 
on large scales via the Sachs--Wolfe effect, 
$O_l\sim 0$ on intermediate
scales ($l\sim 50$ for adiabatic and $l\sim 100$ for isocurvature), 
indicating loss of information on the phases of the underlying fluctuations.
Then, toward smaller scales, we recover information, lose information, 
and so on. 
Exact scales at which $O_l\sim 0$ depend on $r$ and cosmology.
A good news is that a high signal-to-noise map of the CMB polarization 
anisotropy will enable us to overcome the loss of 
information, as the polarization transfer function is out of phases 
in $l$ compared to the temperature transfer function, 
filling up information at which $O_l\sim 0$.
In the other words, the polarization anisotropy has finite information
about the phases of the primordial perturbations, when the temperature 
anisotropy has zero information.

%
% Cubic statistic for the primordial non-Gaussianity
%
%%%%%%%%%%%%%%%%%%%%%%%%%%%%%%%%%%%%%%%%%%%%%%%%%%%%%%%%%%%%%%%%%%
\section{FAST CUBIC STATISTICS}\label{sec:cubic}

\subsection{Primordial Non-Gaussianity}

Using two functions introduced in the previous section, we construct 
a {\it cubic} statistic optimal for the primordial non-Gaussianity.
We apply filters to $a_{lm}$, and then transform the filtered $a_{lm}$'s 
to obtain two maps, $A$ and $B$, given by
%%%%%%%%%%%%%%%%%%%%%%%%%%%%%%%%%%%%%%%%%%%%%%%%%%%%%%%%%%%%%%%%%%%
\begin{eqnarray}
 \label{eq:filter1}
 A(r,\hat{\bm n}) &\equiv&
  \sum_{lm} \frac{\alpha_l(r)b_l}{{\cal C}_l}a_{lm}Y_{lm}(\hat{\bm n}),\\
 \label{eq:filter2}
 B(r,\hat{\bm n}) &\equiv&
  \sum_{lm} \frac{\beta_l(r)b_l}{{\cal C}_l}a_{lm}Y_{lm}(\hat{\bm n}).
\end{eqnarray}
%%%%%%%%%%%%%%%%%%%%%%%%%%%%%%%%%%%%%%%%%%%%%%%%%%%%%%%%%%%%%%%%%%%
The latter map, $B(r,\hat{\bm n})$, is exactly the 
${\cal O}_l$-filtered map, an Wiener-filtered map of the underlying 
primordial fluctuations.
We then form a cubic statistic given by
%%%%%%%%%%%%%%%%%%%%%%%%%%%%%%%%%%%%%%%%%%%%%%%%%%%%%%%%%%%%%%%%%%%
\begin{equation}
 \label{eq:skewness}
  {\cal S}_{prim} \equiv 4\pi \int r^2 dr 
  \int \frac{d^2\hat{\bm n}}{4\pi}
  A(r,\hat{\bm n}) B^2(r,\hat{\bm n}),
\end{equation}
%%%%%%%%%%%%%%%%%%%%%%%%%%%%%%%%%%%%%%%%%%%%%%%%%%%%%%%%%%%%%%%%%%%
where the angular average is done on full sky regardless of sky cut.
We find that ${\cal S}_{prim}$ reduces \textit{exactly} to
%%%%%%%%%%%%%%%%%%%%%%%%%%%%%%%%%%%%%%%%%%%%%%%%%%%%%%%%%%%%%%%%%%%
\begin{equation}
 \label{eq:Sprim}
  {\cal S}_{prim}= \sum_{l_1\le l_2\le l_3}
  \frac{{\cal B}_{l_1l_2l_3}^{obs}{\cal B}_{l_1l_2l_3}^{prim}}
  {{\cal C}_{l_1}{\cal C}_{l_2}{\cal C}_{l_3}},
\end{equation}
%%%%%%%%%%%%%%%%%%%%%%%%%%%%%%%%%%%%%%%%%%%%%%%%%%%%%%%%%%%%%%%%%%%
where 
%%%%%%%%%%%%%%%%%%%%%%%%%%%%%%%%%%%%%%%%%%%%%%%%%%%%%%%%%%%%%%%%%%%
\begin{equation}
{\cal B}_{l_1l_2l_3} \equiv B_{l_1l_2l_3}b_{l_1}b_{l_2}b_{l_3},
\end{equation}
%%%%%%%%%%%%%%%%%%%%%%%%%%%%%%%%%%%%%%%%%%%%%%%%%%%%%%%%%%%%%%%%%%%
and $B_{l_1l_2l_3}^{obs}$ is the observed bispectrum with the effect of $b_l$
corrected while $B_{l_1l_2l_3}^{prim}$ the theoretical one for $f_{NL}=1$ 
\citep{2001PhRvD..63f3002K},
%%%%%%%%%%%%%%%%%%%%%%%%%%%%%%%%%%%%%%%%%%%%%%%%%%%%%%%%%%%%%%%%%%%
\begin{equation}
 \label{eq:primbispectrum}
B^{prim}_{l_1l_2l_3} 
 \equiv 2I_{l_1l_2l_3}\int r^2 dr 
 \left[\beta_{l_1}(r)\beta_{l_2}(r)\alpha_{l_3}(r)
  + \beta_{l_3}(r)\beta_{l_1}(r)\alpha_{l_2}(r)
  + \beta_{l_2}(r)\beta_{l_3}(r)\alpha_{l_1}(r)
 \right],
\end{equation}
%%%%%%%%%%%%%%%%%%%%%%%%%%%%%%%%%%%%%%%%%%%%%%%%%%%%%%%%%%%%%%%%%%%
where 
%%%%%%%%%%%%%%%%%%%%%%%%%%%%%%%%%%%%%%%%%%%%%%%%%%%%%%%%%%%%%%%%%%%
\begin{equation}
 I_{l_1l_2l_3}\equiv 
 \sqrt{\frac{(2l_1+1)(2l_2+1)(2l_3+1)}{4\pi}}
 \left(\begin{array}{ccc}l_1&l_2&l_3\\0&0&0\end{array}\right).
\end{equation}
%%%%%%%%%%%%%%%%%%%%%%%%%%%%%%%%%%%%%%%%%%%%%%%%%%%%%%%%%%%%%%%%%%%
It is straightforward to derive equation~(\ref{eq:Sprim}) from
(\ref{eq:skewness}) using equation~(\ref{eq:primbispectrum}).

The denominator of equation~(\ref{eq:Sprim}) is the variance of 
${\cal B}_{l_1l_2l_3}^{obs}$ in the limit of weak
non-Gaussianity (say $\left|f_{NL}\right|\la 10^3$)
when all $l$'s are different: 
$\left<{\cal B}_{l_1l_2l_3}^2\right>= 
{\cal C}_{l_1}{\cal C}_{l_2}{\cal C}_{l_3}\Delta_{l_1l_2l_3}$,
where $\Delta_{l_1l_2l_3}$ is 6 for $l_1=l_2=l_3$, 2 for 
$l_1=l_2\neq l_3$ etc., and 1 otherwise.
The bispectrum configurations are thus summed up nearly optimally with the 
approximate inverse-variance weights, provided that
$\Delta_{l_1l_2l_3}$ is approximated with $\simeq 1$.
The least-square fit of ${\cal B}_{l_1l_2l_3}^{prim}$ to 
${\cal B}_{l_1l_2l_3}^{obs}$ can be performed to yield
%%%%%%%%%%%%%%%%%%%%%%%%%%%%%%%%%%%%%%%%%%%%%%%%%%%%%%%%%%%%%%%%%%%
\begin{equation}
 \label{eq:Sprim*}
  {\cal S}_{prim} \simeq f_{NL}
  \sum_{l_1\le l_2\le l_3}
  \frac{({\cal B}_{l_1l_2l_3}^{prim})^2}
  {{\cal C}_{l_1}{\cal C}_{l_2}{\cal C}_{l_3}}.
\end{equation}
%%%%%%%%%%%%%%%%%%%%%%%%%%%%%%%%%%%%%%%%%%%%%%%%%%%%%%%%%%%%%%%%%%%
This equation gives an estimate of $f_{NL}$ directly 
from ${\cal S}_{prim}$.

The most time consuming part is the back-and-forth harmonic transformation
necessary for pre-filtering [Eqs.~(\ref{eq:filter1}) and (\ref{eq:filter2})],
taking $N^{3/2}$ operations times the number of sampling points of $r$,
of order 100, for evaluating the integral [Eq.~(\ref{eq:skewness})].
This is much faster than the full bispectrum analysis which
takes $N^{5/2}$, enabling us to perform a more detailed analysis
of the data in a reasonable amount of computational time.
For example, measurements of all bispectrum configurations up to
$l_{max}=512$ take 8 hours to compute on 16 processors of an SGI Origin 
300; thus, even only 100 Monte Carlo simulations take 1 month to carry out.  
On the other hand, ${\cal S}_{prim}$ takes only 30 seconds to compute, 
1000 times faster.
When we measure $f_{NL}$ for $l_{max}=1024$, we speed up by a factor
of 4000: 11 days for the bispectrum vs 4 minutes for ${\cal S}_{prim}$.
We can do 1000 simulations for $l_{max}=1024$ in 3 days.

%
% Correlated isocurvature fluctuations
%
\subsection{Mixed Fluctuations} \label{sec:mix}

The ${\cal O}_l$-filtered map, $B$, is an Wiener-filtered map of
primordial curvature or isocurvature perturbations; however, this is correct 
only when correlations between the two components are negligible.
On the other hand, multi-field inflation models
\citep{Langlois:1999dw,Gordon:2000hv} and curvaton models 
\citep{Lyth:2001nq,Moroi:2001ct} naturally predict correlations.
% ed mixture of adiabatic and isocurvature fluctuations. 
The current CMB data are consistent with, but do not require, a mixture 
of the correlated fluctuations 
\citep{Amendola:2001ni,Trotta:2001yw,2003astro.ph..2225P}.
In this case, the Wiener filter for the primordial fluctuations 
[Eq.~(\ref{eq:Ol})] needs to be modified such that 
${\cal O}_l(r)=\beta_l(r)b_l/{\cal C}_l
\rightarrow \tilde{\beta}_l(r)b_l/{\cal C}_l$, where
%%%%%%%%%%%%%%%%%%%%%%%%%%%%%%%%%%%%%%%%%%%%%%%%%%%%%%%%%%%%%%%%%%%
\begin{eqnarray}
 \label{eq:betaadi_l}
 \nonumber
  \tilde{\beta}^{adi}_l(r) &=& 
  \frac{2}{\pi}\int k^2 dk 
  \left[ P_{\Phi}(k) g_{Tl}^{adi}(k) 
       + P_{C}(k)   g^{iso}_{Tl}(k) \right] j_l(k r), \\
 \label{eq:betaiso_l}
 \nonumber
  \tilde{\beta}^{iso}_l(r) &=& 
  \frac{2}{\pi}\int k^2 dk 
  \left[ P_{S}(k)     g_{Tl}^{iso}(k) 
       + P_{C}(k) g^{adi}_{Tl}(k) \right] j_l(k r),
\end{eqnarray}
%%%%%%%%%%%%%%%%%%%%%%%%%%%%%%%%%%%%%%%%%%%%%%%%%%%%%%%%%%%%%%%%%%%
for curvature ($adi$) and isocurvature ($iso$) perturbations, respectively.
Here $P_{\Phi}$ is the primordial power spectrum of curvature
perturbations, $P_{S}$ of isocurvature perturbations, and $P_{C}$
of cross correlations.

%
% Measuring non-Gaussianity from correlated terms
%
As for measuring non-Gaussianity from the correlated fluctuations,
we use equation~(\ref{eq:phi}) as a model of $\Phi$- and $S$-field
non-Gaussianity to parameterize them with $f_{NL}^{adi}$ and
$f_{NL}^{iso}$, respectively. 
We then form a cubic statistic similar to ${\cal S}_{prim}$ 
[Eq.~(\ref{eq:skewness})], using $A(r,\hat{\bm n})$ and a new filtered 
map $\tilde{B}(r,\hat{\bm n})$ which uses $\tilde{\beta}_l(r)$.
We have two cubic combinations: $A_{adi}\tilde{B}^2_{adi}$ for
measuring $f_{NL}^{adi}$ and $A_{iso}\tilde{B}^2_{iso}$ for
$f_{NL}^{iso}$, each of which comprises four terms including
one $P_{\Phi}^2$ (or $P_{S}^2$), one $P_{C}^2$, 
and two $P_{\Phi}P_{C}$'s (or $P_{S}P_{C}$'s).
In other words, the correlated contribution makes the total number of terms 
contributing to the non-Gaussianity four times more than the 
uncorrelated-fluctuation models (see \citep{Bartolo:2001cw} for more 
generic cases).
%; thus, observational sensitivity to $f_{NL}^{adi}$ or
%$f_{NL}^{iso}$ can be several times better depending on $P_C(k)$
%and $g_{Tl}(k)$.

%
% Cubic statistic for point-source non-Gaussianity
%
\subsection{Point Source Non-Gaussianity}

Next, we show that the filtering method is also useful for 
measuring foreground non-Gaussianity arising from extragalactic 
point sources.
The residual point sources left unsubtracted in a map can 
seriously contaminate both the power spectrum and the bispectrum.
We can, on the other hand, use multi-band observations as well as 
external template maps of dust, free-free, and synchrotron emission, to 
remove diffuse Galactic foreground \citep{2003astro.ph..2208B}.
The radio sources with known positions can be safely masked.

The filtered map for the point sources is 
%%%%%%%%%%%%%%%%%%%%%%%%%%%%%%%%%%%%%%%%%%%%%%%%%%%%%%%%%%%%%%%%%%%
\begin{equation}
 D(\hat{\bm n}) \equiv  
 \sum_{lm} \frac{b_l}{{\cal C}_l} a_{lm} Y_{lm}(\hat{\bm n}).
\end{equation}
%%%%%%%%%%%%%%%%%%%%%%%%%%%%%%%%%%%%%%%%%%%%%%%%%%%%%%%%%%%%%%%%%%%
This filtered map was actually used for detecting point sources
in the {\sl WMAP} maps \citep{2003astro.ph..2208B}.
Using $D(\hat{\bm n})$, the cubic statistic is derived as
%%%%%%%%%%%%%%%%%%%%%%%%%%%%%%%%%%%%%%%%%%%%%%%%%%%%%%%%%%%%%%%%%%%
\begin{equation}
 \label{eq:skewness*}
  {\cal S}_{src} \equiv
  \int \frac{d^2\hat{\bm n}}{4\pi} D^3(\hat{\bm n})
  =
  \frac{3}{2\pi} \sum_{l_1\le l_2\le l_3}
  \frac{{\cal B}_{l_1l_2l_3}^{obs}{\cal B}_{l_1l_2l_3}^{src}}
  {{\cal C}_{l_1}{\cal C}_{l_2}{\cal C}_{l_3}}.
\end{equation}
%%%%%%%%%%%%%%%%%%%%%%%%%%%%%%%%%%%%%%%%%%%%%%%%%%%%%%%%%%%%%%%%%%%
Here, $B_{l_1l_2l_3}^{src}$ is the point-source bispectrum for unit
white-noise bispectrum,
%%%%%%%%%%%%%%%%%%%%%%%%%%%%%%%%%%%%%%%%%%%%%%%%%%%%%%%%%%%%%%%%%%%
%\begin{equation}
% \label{eq:psbispectrum}
  $B^{src}_{l_1l_2l_3} = I_{l_1l_2l_3}$.
%\end{equation}
%%%%%%%%%%%%%%%%%%%%%%%%%%%%%%%%%%%%%%%%%%%%%%%%%%%%%%%%%%%%%%%%%%%
($b_{src}=1$ in \citet{2001PhRvD..63f3002K}.)
When covariance between $B_{l_1l_2l_3}^{prim}$ and
$B_{l_1l_2l_3}^{src}$ is negligible as is the case for {\sl WMAP} and 
{\sl Planck} \citep{2001PhRvD..63f3002K}, we find
%%%%%%%%%%%%%%%%%%%%%%%%%%%%%%%%%%%%%%%%%%%%%%%%%%%%%%%%%%%%%%%%%%%
\begin{equation}
 \label{eq:Sps}
  {\cal S}_{src} \simeq
  \frac{3b_{src}}{2\pi} 
  \sum_{l_1\le l_2\le l_3}
   \frac{({\cal B}_{l_1l_2l_3}^{src})^2}
   {{\cal C}_{l_1}{\cal C}_{l_2}{\cal C}_{l_3}}.
\end{equation}
%%%%%%%%%%%%%%%%%%%%%%%%%%%%%%%%%%%%%%%%%%%%%%%%%%%%%%%%%%%%%%%%%%%
We omit the covariance only for simplicity; including
it is simple \citep{2001PhRvD..63f3002K,2002ApJ...566...19K}.
% When we measure $a_{lm}^{obs}$ from equation~(\ref{eq:alm_obs}),
% we divide ${\cal S}_{src}$ by $\int d^2\hat{\bm n}M^3(\hat{\bm n})/(4\pi)$.
Again ${\cal S}_{src}$ measures $b_{src}$ much faster than the full
bispectrum analysis, constraining effects of residual point sources
on CMB sky maps. Since ${\cal S}_{src}$ does not contain the extra integral
over $r$, it is even 100 times faster to compute than ${\cal S}_{prim}$.
This statistic is particularly useful because it is sometimes difficult to
tell how much of $C_l$ is due to point sources.
\citet{2003astro.ph..2223K} have used ${\cal S}_{src}$ 
(i.e., $b_{src}$) to measure $C_l$ due to the unsubtracted point sources.

%
% observed a_lm
%
\subsection{Incomplete Sky Coverage}

Finally, we show how to incorporate incomplete sky coverage and
pixel weights into our statistics. 
Suppose that we weight a sky map by $M(\hat{\bm n})$ to measure 
the harmonic coefficients,
%%%%%%%%%%%%%%%%%%%%%%%%%%%%%%%%%%%%%%%%%%%%%%%%%%%%%%%%%%%%%%%%%%%
\begin{equation}
a_{lm}^{obs}= T^{-1}\int d^2\hat{\bm n} M(\hat{\bm n})
\Delta T(\hat{\bm n}) Y_{lm}^*(\hat{\bm n}).
\end{equation}
%%%%%%%%%%%%%%%%%%%%%%%%%%%%%%%%%%%%%%%%%%%%%%%%%%%%%%%%%%%%%%%%%%%
A full-sky $a_{lm}$ is related to $a_{lm}^{obs}$ 
through the coupling matrix 
$M_{ll'mm'}\equiv \int d^2\hat{\bm n} M(\hat{\bm n})
Y_{lm}^*(\hat{\bm n})Y_{l'm'}(\hat{\bm n})$ by
$a_{lm}^{obs}=\sum_{l'm'}a_{l'm'}M_{ll'mm'}$. 
In this case the observed bispectrum is biased by a factor of
$\int d^2\hat{\bm n}M^3(\hat{\bm n})/(4\pi)$;
thus, we need to divide $S_{prim}$ and $S_{src}$ by this factor.
If only sky cut is considered, then this factor is a fraction of 
the sky covered by observations \citep{2002ApJ...566...19K}.

We have carried out extensive Monte Carlo simulations of non-Gaussian sky maps 
computed with equation~(\ref{eq:alm}).
Appendix A of \citet{2003astro.ph..2223K} describes the simulations in detail.
We find that ${\cal S}_{prim}$
reproduces input $f_{NL}$'s accurately both on full sky and 
incomplete sky with modest Galactic cut and inhomogeneous noise
expected for {\sl WMAP}, i.e., the statistic is unbiased.
We cannot however make a sky cut very large, 
e.g., more than 50\% of the sky, as for which the covariance matrix of 
${\cal B}_{l_1l_2l_3}$ is no longer diagonal.
The cubic statistic does not include the off-diagonal 
terms of the covariance matrix [see Eq.~(\ref{eq:Sprim})];
however, it works fine for {\sl WMAP} sky maps for which we can use
more than 75\% of the sky.
Also, we have found that equation~(\ref{eq:Sps}) correctly
estimates $b_{src}$ using simulated realizations of point sources
(see Appendix B of \citet{2003astro.ph..2223K}).
As for uncertainty in the cubic statistics, 
Figure~\ref{fig:error} shows that errors of $f_{NL}$ from ${\cal S}_{prim}$ 
and $b_{src}$ from ${\cal S}_{src}$ are as small as those from the full 
bispectrum analysis (see descriptions in Appendix).

%
% Summary
%
%%%%%%%%%%%%%%%%%%%%%%%%%%%%%%%%%%%%%%%%%%%%%%%%%%%%%%%%%%%%%%%%%%
\section{CONCLUSION}\label{sec:conclusion}

Using the method described in this paper, we can measure non-Gaussian 
fluctuations in a nearly full-sky CMB map much faster than the 
full bispectrum 
analysis without loss of sensitivity (see Appendix).
Our fast statistics allow us to carry out extensive Monte Carlo
simulations characterizing effects of realistic noise properties of 
experiments, sky cut, foreground sources, and so on.  
A reconstructed map of the primordial fluctuations, which 
plays a key role in our method, potentially gives other real-space 
statistics more sensitivity to primordial non-Gaussianity.
As we have shown our method can be applied to the primordial 
non-Gaussianity arising from inflation, gravity, or 
correlated isocurvature fluctuations, as well as the foreground 
non-Gaussianity from radio point sources, all of which can be important 
sources of non-Gaussian fluctuations on the forthcoming CMB sky maps.

%
% Appendix
%
%%%%%%%%%%%%%%%%%%%%%%%%%%%%%%%%%%%%%%%%%%%%%%%%%%%%%%%%%%%%%%%%%%
\appendix
\section{PERFORMANCE OF CUBIC STATISTICS}
\label{sec:performance}

\begin{figure}
% \leavevmode \epsfxsize=8.5cm \epsfbox{figure1.eps}
% \includegraphics{figure1.eps}% Here is how to import EPS art
 \plotone{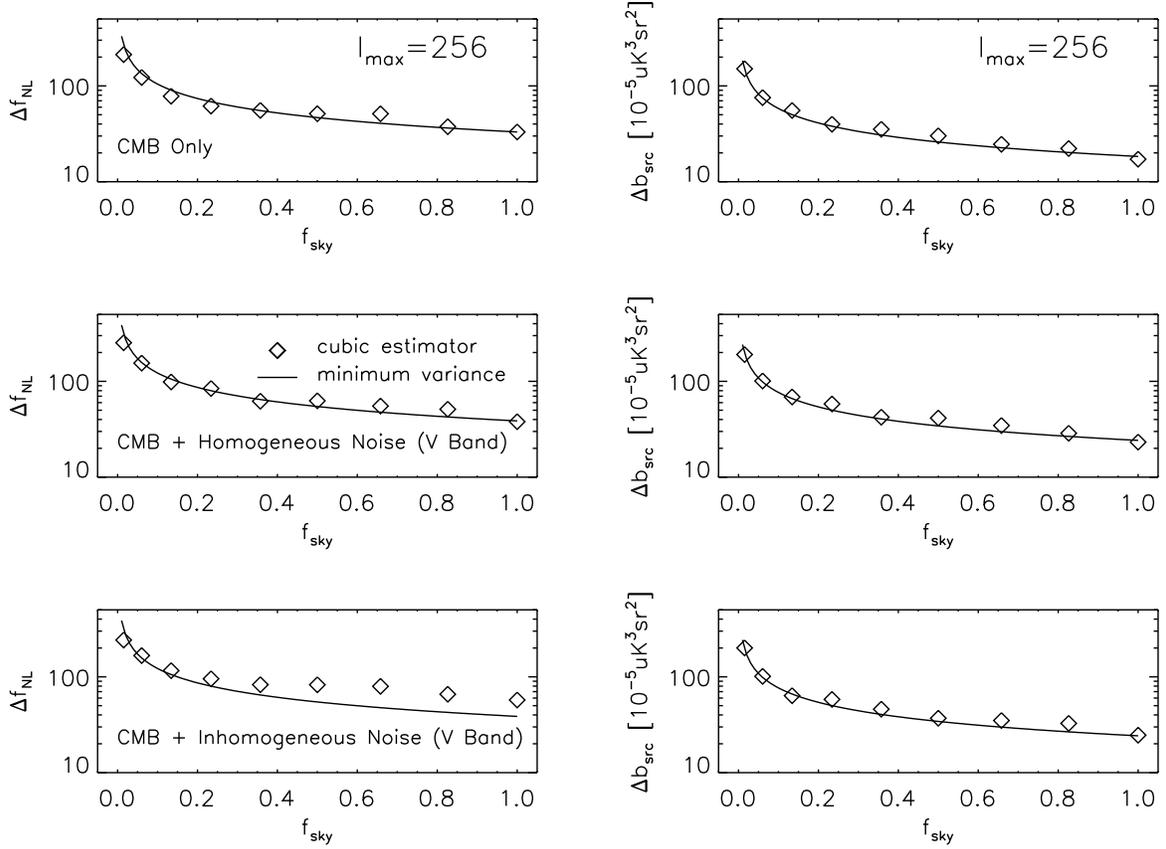}
 \caption{\label{fig:error} 
 Performance of cubic statistics. The left panels and right panels
show errors of $f_{NL}$ and $b_{src}$, respectively, which are 
obtained from
300 Gaussian simulations. Each point has been computed for
a given straight sky cut with $|b|_{cut}$.  
From the right to left, $|b|_{cut}=0$, 10,
20, 30, 40, 50, 60, 70, and $80^\circ$.
The solid lines show the minimum variance which would be obtained
by the full bispectrum analysis. 
The top, middle, and bottom panels show simulations of the CMB signal only, 
CMB plus homogeneous noise, and CMB plus inhomogeneous noise, respectively.
Noise properties assume WMAP 1-year data in V band.
}
\end{figure}

We have extensively tested our cubic statistics using
Monte Carlo simulations of CMB sky with realistic noise properties
and various galaxy masks.
More specifically, we have used {\sl WMAP} 1-year noise properties
in V band \citep{2003astro.ph..2207B} and
straight masks in Galactic coordinates
with $|b|_{cut}=0$, 10, 20, 30, 40, 50, 60, 70, and $80^\circ$.
(I.e., $-|b|_{cut}<b<|b|_{cut}$ has been masked.)
Figure~\ref{fig:error} shows uncertainty in $f_{NL}$ and
$b_{src}$ obtained from 300 
Gaussian simulations using the cubic statistics (diamonds)
for different galaxy masks. 
The solid lines show the minimum variance which would be expected
for the full bispectrum analysis. These lines have been computed by
$\sqrt{F^{-1}_{ii}/f_{sky}}$, where $F_{ij}$ is the Fisher matrix
\citep{2001PhRvD..63f3002K} and $f_{sky}$ is a fraction of sky
surviving the mask.
We find that the cubic statistics perform remarkably well, 
giving errors as small as the full bispectrum analysis for
all masks, for simulations of the CMB signal only and CMB with 
homogeneous noise.
(The homogeneous noise has r.m.s. noise which matches the average
noise in V band.)
However, the statistics perform a bit worse in the presence of 
inhomogeneous noise -- this is because we are using a simple 
uniform weighting ($M(\hat{\bm n})=0$ for the masked pixels and
$M(\hat{\bm n})=1$ otherwise)
rather than the optimal $C^{-1}$ weighting.
Since noise in some pixels are much larger than average,
our statistics are affected by those noisier pixels. 
It is not straightforward to implement $C^{-1}$ weighting 
(which is non-diagonal) in our cubic statistics, but
we continue to investigate a way to take into account
inhomogeneous noise in our method.

%%%%%%%%%%%%%%%%%%%%%%%%%%%%%%%%%%%%%%%%%%%%%%%%%%%%%%%%%%%%%%%%%%
%%%%%%%%%%%%%%%%%%%%%%%%%%%%%%%%%%%%%%%%%%%%%%%%%%%%%%%%%%%%%%%%%%

%%%%%%%%%%%%%%%%%%%%%%%%%%%%%%%%%%%%%%%%%%%%%%%%%%%%%%%%%%%%%%%%%% 
% \bibliography{nongaus}% Produces the bibliography via BibTeX.
% \bibliographystyle{apj}
% \bibliography{apj-jour,nongaus}

\end{document}